\documentclass[slac_one]{revtex4}
\usepackage{graphicx}
\usepackage{fancyhdr}
\begin{document}

\title{Heavy b' and t' Decay} 

%

\author{Yuan Chao, Kai-Feng Chen, Shing-Kuo Chen, George W.S. Hou, Bo-Yan Huang, Yeong-Jyi Lei}
\affiliation{National Taiwan University}
%

\begin{abstract}
Heavy $t' \to bW$ is currently being searched for at the Tevatron, 
but a broader spectrum should be explored at the LHC. For $m_{b'} < m_{t'}$ , 
we discuss the two decay branches, $b' \to tW^*$ and $t^*W$, below the $tW$ threshold, 
and how they merge with $b' \to tW$ above threshold. We use a
convolution method with five body final state, and compare the 
production cross section with PYTHIA. This can be extended for a $b'$ with
mass heavier than $t'$. A similar discussion is given for $t'$ decay, 
considering $t' \to b^{(*)}W^{(*)}$ processes, besides the $t' \to bW$ considered by CDF.
\end{abstract}

\maketitle

\thispagestyle{fancy}


\section{INTRODUCTION} 

Three generation of quarks was first proposed by Kobayashi and Maskawa in 1960s 
in order to provide the irreducible phase in the quark mixing matrix and carry out 
the existance of $CP$ violation. 
Such picture becomes the basis of the Standard Model (SM),
explains many physics phenomena, and fits the experimental data pretty well.
However, SM is not yet a complete theory and the options for new physics phenomena are still open.
The fourth generation of quarks may still exist within 
the SM framework~\cite{Arhrib:2000ct,Hou:2005fp}.
Through out this report, we use the notations of $t^\prime$ and $b^\prime$ represent the 
up- and down-type fourth generation of quarks, respectively.

The CDF and D0 are the energy frontier experiments at the Tevatron.
They utilized their proton--anti-proton beams at a center-of-mass energy of 1.96 TeV to 
perform direct searches for the fourth generation quarks. 
The best limit on the $b^\prime$ mass ($>268$ GeV/$c^2$ at 95\% confidence intervals) is 
provided by CDF using the $Z+$jets events~\cite{Scott:2006}. 
However, a 100\% branching fraction of $b^\prime \to bZ$ decay is assumed in the analysis.
Since $b^\prime \to bZ$ is a FCNC decay, the branching fraction should be small and the 
limit on $b^\prime$ mass is not really solid. 
For the search of a top-like heavy quark, CDF gives a mass limit of 311 GeV/$c^2$ 
at 95\% confidence level for the $t^\prime \to qW$ final state. This study will be 
continued at the LHC experiments. 

In this report, the decay branches $b^\prime \to t^{(*)}W^{(*)}$ are studied
using a convolution method with a five body final state at tree level.
If the mass of $b^\prime$ is below the $tW$ threshold, 
the decay branches, $b^\prime \to tW^*$ and $t^*W$  are expected to be dominant.
We analyze how these decay branches merge with $b' \to tW$ decays when $m_{b^\prime}$ above the $tW$ threshold. 

\begin{figure*}[h]
\centering
\includegraphics[width=60mm,height=40mm]{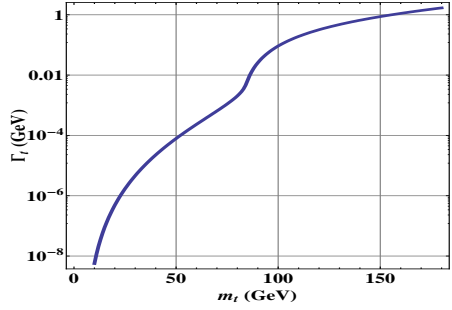}
\caption{Decay width of $t \to bW$. A clear effect of the $W$ mass threshold is seen.} \label{tbwwid}
\end{figure*}

The decay width for a top quark is rather simple. 
As shown in figure~\ref{tbwwid}, the decay width of top assuming 100\%
branching fraction to a bottom quark and a $W$-boson is calculated~\cite{Calderon:2001qq,BarShalom:2005cf}.
The width of $W$-boson is considered in the estimation:
\begin{eqnarray*}\label{eq:ttobw}
m_t &>& m_W + m_b \Rightarrow t \to bW~,	  \\ 
m_t &<& m_W + m_b \Rightarrow t \to bW^*~.    \\
\end{eqnarray*}
A threshold effect for the width around 80 $GeV$ can be seem. 
In the calculation for the fourth generation $b' \to tW$ decays, 
the width of top should be considered as well~\cite{Arhrib:2000ct}.

Different assumptions of the magnitude of $V_{cb'}$ and $V_{tb'}$
will lead to different dominant $b'$ decay channels, 
if $m_{b'}$ is less than $m_t + m_W \approx 255$ GeV/$c^2$. 
The FCNC decay, $b^\prime \to bZ$, is also possible but it is suppressed 
by the second order diagram. 
In the case of  $m_{b'} > 255$ GeV/$c^2$, the decay channel $b^\prime \to tW$ is open and 
is expected to be dominant. The transition $b'\to cW$ would be suppressed 
for such a heavy $b^\prime$ scenario except in a scenario with a 
large value of $V_{cb'}$.
If we only consider the $t^{(*)}W^{(*)}$ final state, the following 
decay signature is expected:
\begin{eqnarray*}\label{eq:bptotw}
m_{b^\prime} &>& m_t + m_W \Rightarrow b' \to tW~,      \\ 
m_{b^\prime} &<& m_t + m_W \Rightarrow b' \to tW^*(t^*W)~.    \\
\end{eqnarray*}

The decay branches of $t^\prime$ is also interesting. In principle 
the decay $t^\prime \to bW$ is one of the major channels  
due to the electroweak precision test which gives the constraint $|m_{t^\prime}-m_{b^\prime} < m_W|$.
However, $\Gamma(t^\prime \to b'W) > \Gamma(t^\prime \to bW)$ is still possible with 
large  $|V_{t^\prime b^\prime}|$ but with relatively small $|V_{t^\prime b}|$.

\section{$b^\prime \to tW$ DECAY WIDTH}

\begin{figure*}[h]
\centering
\includegraphics[width=60mm,height=40mm]{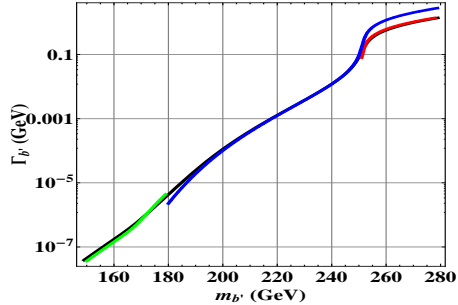}
\caption{The decay width of $b^\prime \to tW$. The solid black curve shows 
the result from a full five-body analysis. The blue curve is the approximation with 
a three-body assumption, and has a double-counting effect when $m_{b^\prime}$ is above 
the $tW$ mass threshold. The red and green curves represent 
another two simplified model assuming two- and four-body decays.} \label{bptwwid}
\end{figure*}

The decay $b^\prime \to tW \to bWW \to bf_if_jf_kf_l$ is treated as a five-body decay. 
The decay width for tree level diagrams can be estimated with a convolution 
method. The result is shown by the solid black curve in Figure~\ref{bptwwid}. 
In the $b'$ mass range of 180--255 GeV$/c^2$,
the decay can be approximated with a three-body model. Basically the 
total decay width of $b^\prime$ is contributed by two sub-processes:
$b^\prime \to t^*W$ and $tW^*$ (either $t$ or $W$ is off-shell).
Such approximation starts to deviate from the best evaluation of five-body
if the mass of $b^\prime$ is above the $tW$ mass threshold. 
It is clear that the width is over-estimated by a factor of two since 
the two processes, $b^\prime \to t^*W$ and $tW^*$, become exactly the same process for heavy $b^\prime$.
Such trouble can be resolved with another two-body approximation of $b^\prime \to tW$ decay.
If the mass of $b^\prime$ is even lower then 180 GeV/$c^2$, then it will force to have another 
off-shell $W$-boson in the final state. In such case, a four-body model can be used in the 
calculation and it is consistent with the full five-body analysis.

\section{THE DECAY OF $b^\prime$ AND $t^\prime$ AND CKM FACTORS}

In this section we discuss the branching ratios for $t^\prime$ and $b^\prime$
and their relationship with the CKM elements. All the amplitudes are 
calculated up to the tree level.

\begin{figure*}[h]
\centering
\includegraphics[width=60mm,height=40mm]{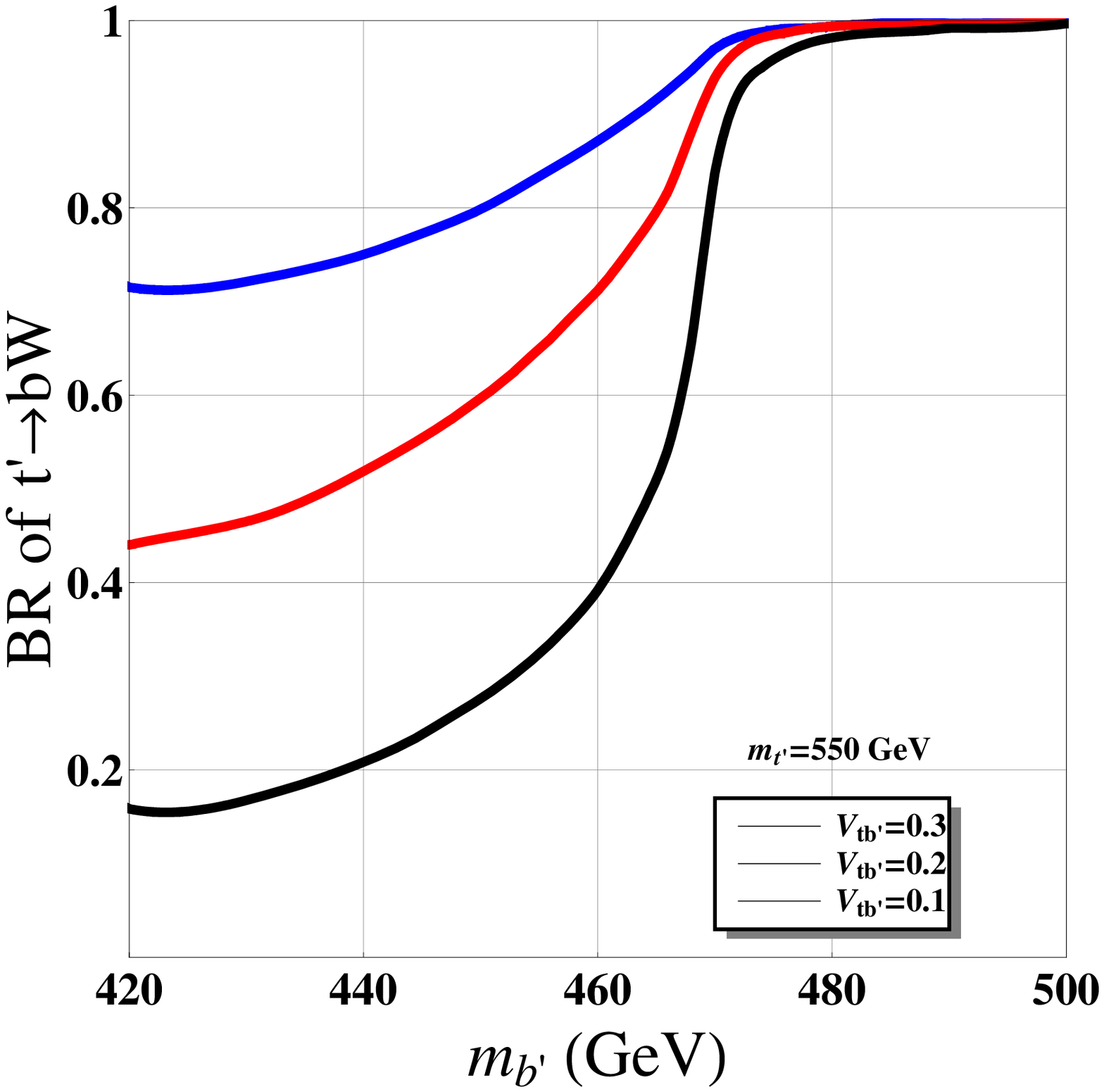}
\includegraphics[width=60mm,height=40mm]{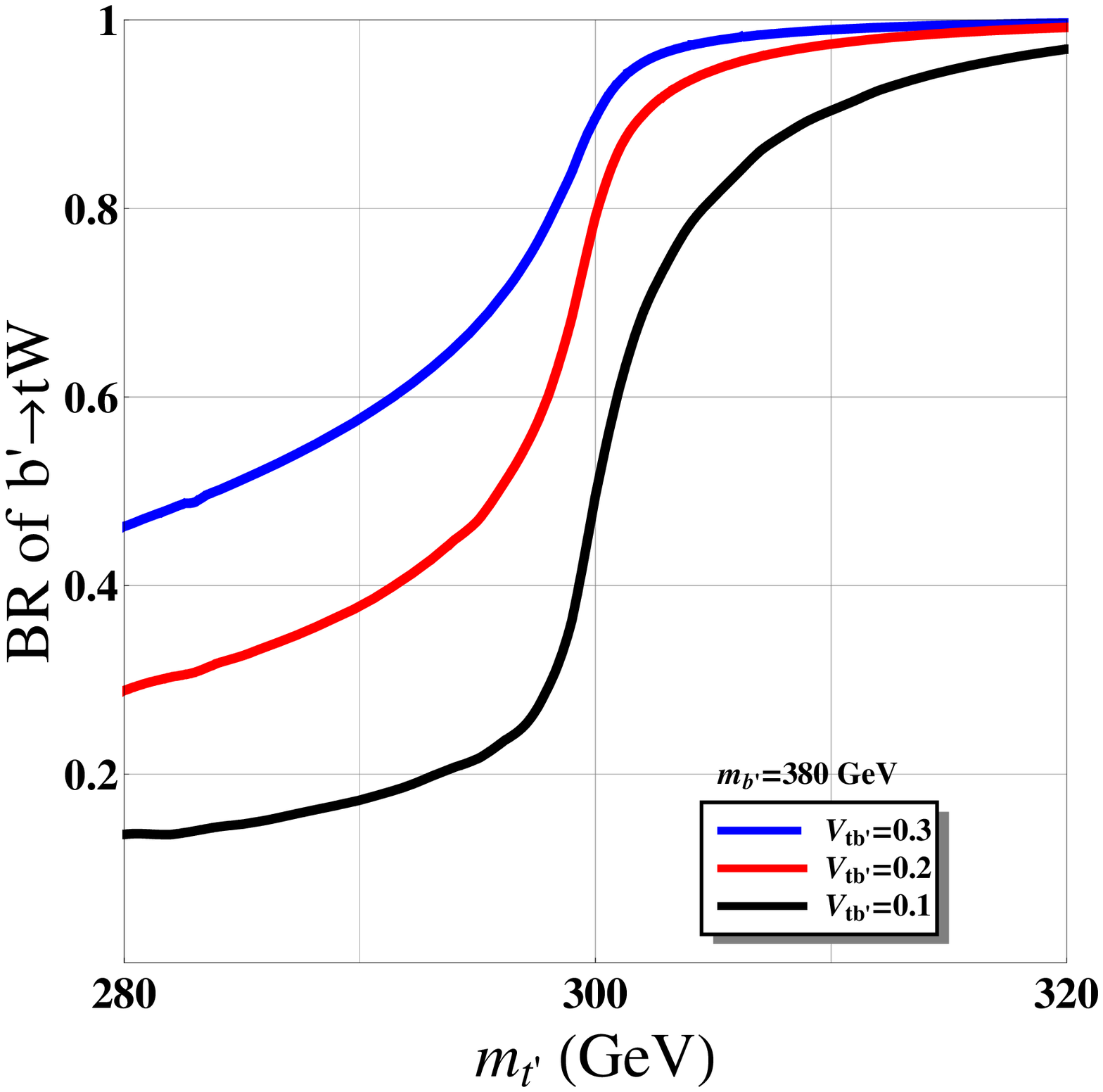}
\caption{Branching fractions for $t' \to b'W$ (left) and $b' \to t'W$ (right) decays.
The different curves represent different magnitudes of $|V_{t'b}|$ and $|V_{tb'}|$. } \label{bp380}
\end{figure*}

The branch ratio of $t' \to bW$ as a function of the mass of $b^\prime$
is shown in Figure~\ref{bp380}, while The mass of $t^\prime$ is fixed at 550 GeV/$c^2$. 
Three different values of $|V_{t'b}|$ are considered as shown in the same figure. 
The curve at the bottom stands for $|V_{t'b}|=0.1$, the one at middle ploted with $|V_{t'b}|=0.2$,
and the upper one indicates for the result with $|V_{t'b}|=0.3$.
The curves can be took as a balance between CKM and phase space. $t' \to b'W$ is
favored by CKM but suppressed by phase space, and on the other hand, $t' \to bW$
is favored by phase space but suppressed by the smaller CKM element. The
plots also show that the branch ratio of $t' \to bW$ can get diluted by
$t' \to b'W$ if the $m_{t'}-m_{b'} < m_{W}$ constrain from electroweak 
precision test (EWPr) does not hold and $m_{b'} < m_{t'} - m_{W}$.

Figure~\ref{bp380} also shows the branch ratios of $b' \to tW$ respecting
to the mass of $t'$, assuming $m_{t'}$ is smaller then the mass
of $b'$. Analogously, the dependence on the CKM factor is also examined. 
The result with $|V_{tb'}|=0.3$ is shown as the upper curve in the figure,
while $|V_{tb'}|=0.3$ for the middle one and $|V_{tb'}|=0.1$ for the buttom one. 
The effects between CKM element and phase space also occur in the branching fractions 
of $b' \to tW$. The decay $b' \to t'W$ is dominant if the EWPr 
constrain does not hold and $m_{t'} < m_{b'} - m_{W}$.

\section{A COMPARISON WITH PYTHIA}

\begin{figure*}[h]
\centering
\includegraphics[width=60mm,height=40mm]{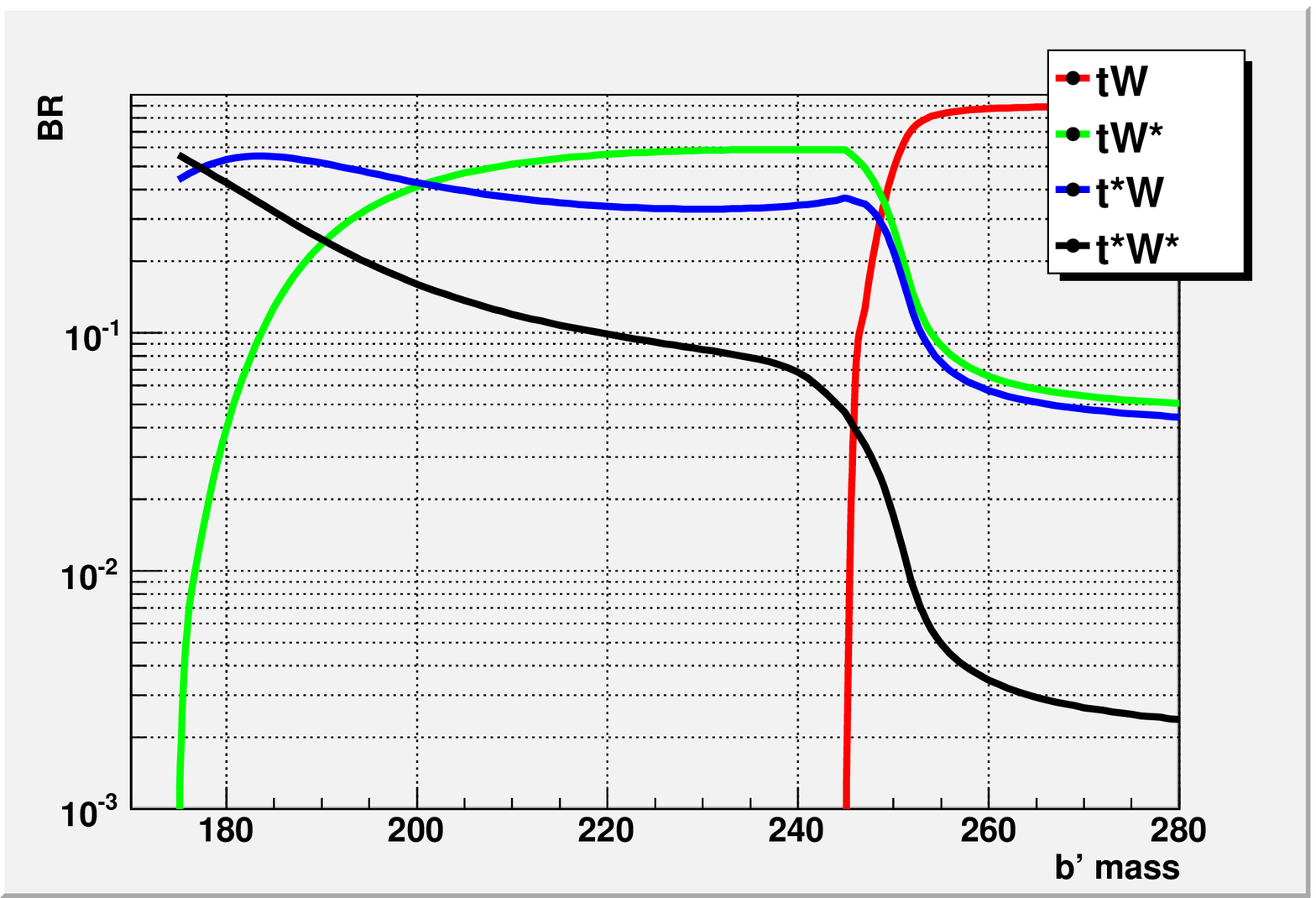}
\includegraphics[width=60mm,height=40mm]{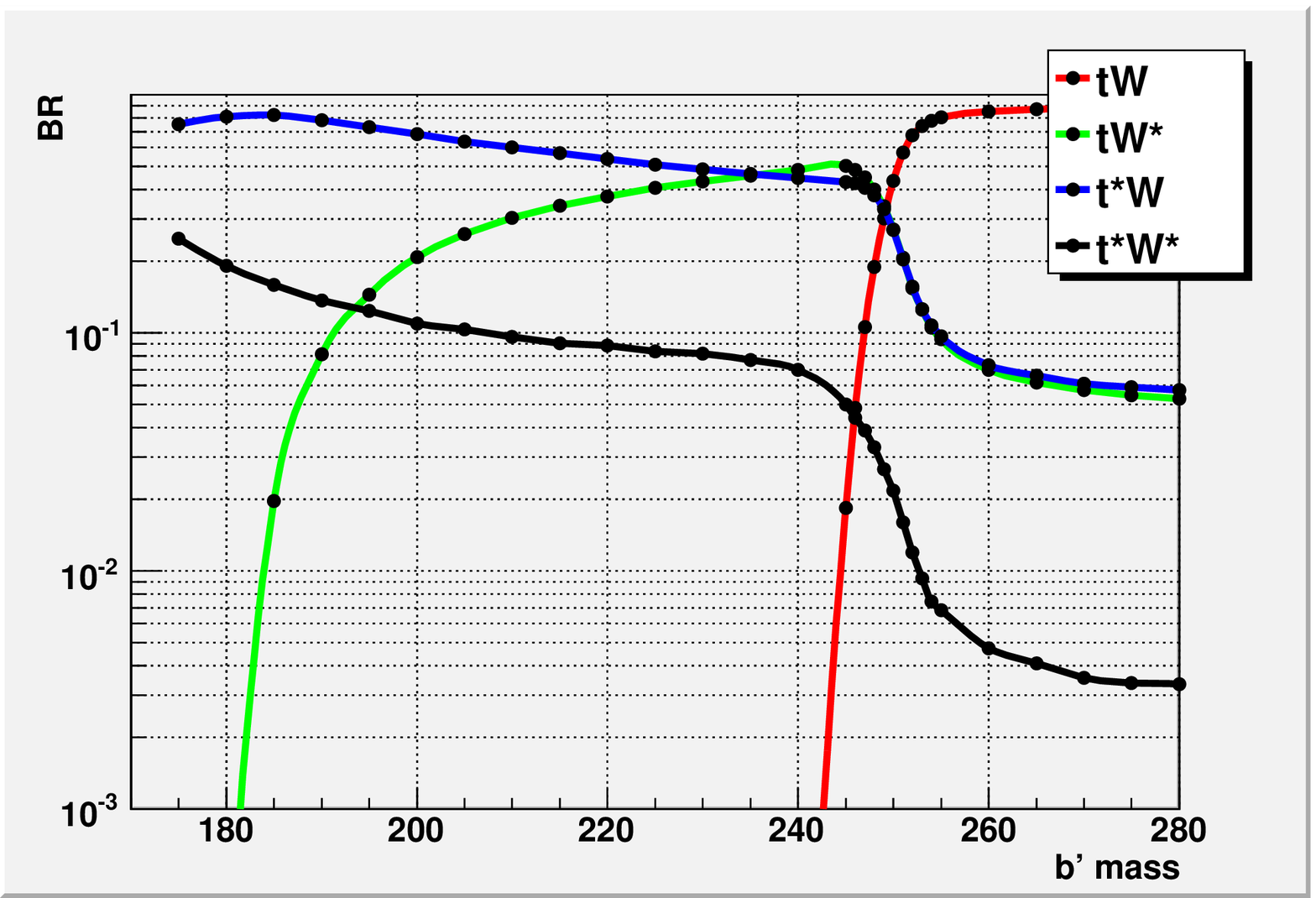}
\caption{The branching ratios of $b' \to tW$, $t^*W$, $tW^*$, and $t^*W^*$ decays.
The left plot is the results from our five-body treatment. The right plot is the results from PYTHIA.} \label{pythia}
\end{figure*}

Finally, we compare the decay rate of $b^\prime \to t^{(*)}W^{(*)}$ from 
our five-body calculations and the results generated by PYTHIA 6~\cite{Sjostrand:2006za}.
The results are shown in Figure~\ref{pythia}. In this comparison,
if the mass of top or $W$ is smaller than its nominal mass by $3 \times \Gamma$,
then these particles are treated as virtual particles. The $\Gamma$
is the width of top or $W$. 
It is clear that our exclusive calculations differ from the results from PYTHIA. 
The source of this deviation is still on inspection.

\begin{thebibliography}{9}   

\bibitem{Scott:2006}
 CDF Collaboration, A.~L. Scott and D.~Stuart, 
 ``Search for New Particles Decaying to $Z^0$+jets,'' 
 {\em CDF Note} {\bf 8590} (2006).


\bibitem{Hou:2005fp}
 W.~S.~Hou and A.~Soddu,
 ``eV seesaw with four generations,''
 Phys.\ Lett.\  B {\bf 638}, 229 (2006)
 [arXiv:hep-ph/0512278].

\bibitem{BarShalom:2005cf}
 S.~Bar-Shalom, G.~Eilam, M.~Frank and I.~Turan,
 ``Width effects on near threshold decays of the top quark t --> c W W, c  Z Z
 and of neutral Higgs bosons,''
 Phys.\ Rev.\  D {\bf 72}, 055018 (2005)
 [arXiv:hep-ph/0506167].

\bibitem{Arhrib:2000ct}
 A.~Arhrib and W.~S.~Hou,
 ``Window on Higgs boson: Fourth generation b' decays revisited,''
 Phys.\ Rev.\  D {\bf 64}, 073016 (2001)
 [arXiv:hep-ph/0012027].

\bibitem{Calderon:2001qq}
 G.~Calderon and G.~Lopez Castro,
 ``Finite W boson width effects in the top quark width,''
 arXiv:hep-ph/0108088.

\bibitem{Sjostrand:2006za}
  T.~Sjostrand, S.~Mrenna and P.~Skands,
  ``PYTHIA 6.4 physics and manual,''
  JHEP {\bf 0605}, 026 (2006)
  [arXiv:hep-ph/0603175].

\end{thebibliography}
\end{document}